\documentclass[twocolumn,floatfix,pra,showpacs,preprintnumbers,superscriptaddress,amsmath,amssymb,10pt,aps,prl]{revtex4}

\usepackage{amsmath}
\usepackage[export]{adjustbox}
\usepackage{graphicx,subfigure,amsmath,bbm}
\usepackage{mathtools}
\usepackage{mathptmx}
\usepackage{subfigure}
\usepackage{psfrag,graphicx}
\usepackage{dcolumn}
\usepackage{amsmath,amssymb}
\usepackage{bm}
\usepackage{color}
\usepackage{latexsym}
\usepackage{epstopdf}
\usepackage{color}
\usepackage[english]{babel}
\usepackage{latexsym}
\usepackage{psfrag,graphicx}
\usepackage{subfigure}
\usepackage{amsmath}
\usepackage{amssymb}
\usepackage{amsfonts}
\usepackage{bm}
\usepackage{natbib}
\usepackage{epstopdf}
\usepackage{xcolor}
\usepackage{leftindex}
\usepackage{appendix}
\newcommand{\RNum}[1]{\uppercase\expandafter{\romannumeral #1\relax}}

\definecolor{mygrey}{gray}{0.35}
\definecolor{myblue}{rgb}{0.2,0.2,0.8}
\definecolor{myzard}{cmyk}{0,0,0.05,0}
\definecolor{mywhite}{rgb}{1,1,1}
\definecolor{mywhite}{rgb}{1,1,1}
\definecolor{myred}{rgb}{1,0.,0.3}

\usepackage[colorlinks=true,citecolor=myblue,linkcolor=myred]{hyperref}

\def\ba{\begin{align}}
\def\enda{\end{align}}
\def\bi{\begin{itemize}}
\def\ei{\end{itemize}}

\def\be{\begin{equation}}
\def\ee{\end{equation}}
\def\bea{\begin{eqnarray}}
\def\eea{\end{eqnarray}}
\def\bse{\begin{subequations}}
\def\ese{\end{subequations}}

\begin{document}
\title{Super-Heisenberg scaling of the quantum Fisher information using spin-motion states}
\def\correspondingauthor{\footnote{Corresponding author: pivanov@phys.uni-sofia.bg}}
\author{Venelin P. Pavlov}
\affiliation{Center for Quantum Technologies, Department of Physics, St. Kliment Ohridski University of Sofia, James Bourchier 5 blvd, 1164 Sofia, Bulgaria}
\author{Peter A. Ivanov}
\affiliation{Center for Quantum Technologies, Department of Physics, St. Kliment Ohridski University of Sofia, James Bourchier 5 blvd, 1164 Sofia, Bulgaria}

\begin{abstract}
We propose a spin-motion state for high-precision quantum metrology with super-Heisenberg scaling of the parameter estimation uncertainty using a trapped ion system. Such a highly entangled state can be created using the Tavis-Cummings Hamiltonian which describes the interaction between a collective spin system and a single vibrational mode. Our method relies on an adiabatic evolution in which the initial motional squeezing is adiabatically transferred into collective spin squeezing. In the weak squeezing regime, we show that the adiabatic evolution creates a spin-squeezed state, which reduces the quantum projective noise to a sub-shot noise limit. For strong bosonic squeezing we find that the quantum Fisher information follows a super-Heisenberg scaling law $\propto N^{5/2}$ in terms of the number of ions $N$. Furthermore, we discuss the spin squeezing parameter which quantifies the phase sensitivity enhancement in Ramsey spectroscopic measurements and show that it also exhibits a super-Heisenberg scaling with $N$. Our work enables the development of high-precision quantum metrology based on entangled spin-boson states that lead to faster scaling of the parameter estimation uncertainty with the number of spins.   

\end{abstract}

\maketitle

\emph{Introduction.-} 
Quantum metrology is one of the most promising branches of quantum technologies that aims to use entanglement to improve measurement precision \cite{Degen2017,Pezze2018,Giovannetti2011,Giovannetti2006}. The sensitivity of the parameter estimation is closely related to the initial state of the quantum probe system. A well known example in precise metrology is the Ramsey interferometer which uses $N$ uncorrelated two-state systems to measure a parameter $\varphi$ with statistical uncertainty $\delta\varphi\propto N^{-1/2}$ which is known as shot-noise limit (SNL). The quantum correlations may yield a favorable scaling of the precise measurements, which overcome the SNL. Indeed, using nonclassical spin squeezed states as a resource for quantum metrology can lead to a reduction in the quantum projective noise which improves the sensitivity to the sub shot-noise limit where the scaling law of the statistical uncertainty obeys $\delta\varphi\propto N^{-l}$ with $l>1/2$ \cite{Kitagawa1993,Ma2011}. Exploring stronger entangled states may further improve the sensitivity to the so-called Heisenberg scaling (limit) $\delta\varphi_{\rm H}\propto N^{-1}$. Over the past few years the creation of spin squeezed states has attracted considerable theoretical \cite{Comprain2022,Comparin2022_1,Swan2024,Carrasco2024,Feng2024,Pavlov2023,Perlin2020} and experimental \cite{Bonhet2016,Hosten2016,Penafiel2020,Huang2023,Bornet2023,Franke2023} interest in the light of their applications for generation of entangled-enhanced quantum sensors.


In this work we propose an entangled quantum state between a collective spin system and a quantum harmonic oscillator for high precision quantum metrology with \emph{super-Heisenberg} scaling of the parameter estimation precision. Such a scaling can be achieved, for example, by utilizing a $k$-body interaction between the spins which enhances the phase sensitivity to $\delta\varphi_{\rm SH}\propto N^{-k}$ with $k>1$ \cite{Boixo2007,Choi2008,Roy2008,Napolitano2011,Zwierz2012}, or by using critical quantum systems \cite{Rams2018,Gietka2023,Lyu2020}. Instead of using such approaches that require a strong interaction between the particles, here we show that the spin-boson entanglement can lead to a super-Heisenberg estimation precision scaling. The preparation of our entangled spin-boson state relies on using the well-known Tavis-Cummings model in quantum optics which describes the interaction between an ensemble of spin-$1/2$ particles and a single bosonic mode. We show that the squeezing in the initial bosonic state of the harmonic oscillator can be adiabatically transferred to the collective spin states. We find two regimes depending on the initial bosonic squeezing. In the regime of weak squeezing the adiabatic transition creates a spin-squeezed state which is disentangled with the harmonic oscillator. We show that such a state can be used for precise quantum metrology with sub shot-noise limited sensitivity. Remarkably, we define a regime of strong bosonic squeezing where the adiabatic transition gives rise to entanglement between the collective spin states and a large set of the bosonic Fock states. We analytically show and numerically confirm that the quantum Fisher information (QFI), a measure which quantifies the sensitivity of the parameter estimation, exhibits a super-Heisenberg scaling law $\propto N^{5/2}$. Hence, the ultimate phase sensitivity which is inversely proportional to the square root of the QFI follows the scaling law $\delta\varphi=\gamma N^{-5/4}$ where $\gamma$ is a constant. Furthermore, we show that the spin squeezing parameter which quantifies the sensitivity enhancement for Ramsey spectroscopic measurement \cite{Wineland1992} also reveals super-Heisenberg scaling. The very strong initial bosonic squeezing, however, suppresses the  
further phase sensitivity enhancement in the sense that the metrological gain is always bounded by the Heisenberg limit.  
 
We show that the proposed spin-boson entangled state can be realized in a trapped ions setup. Here a spin-1/2 is encoded into a pair of internal states $\left|\uparrow\right\rangle$ and $\left|\downarrow\right\rangle$ for each ion. The bosonic degree of freedom is represented by the quantized vibrational mode of the ion chain. The ability to control the motion and the internal states of the trapped ions with high accuracy allows one to manipulate and couple spin and motional degrees of freedom \cite{Wineland1998,Haffner2008,Schneider2012,Monroe2021}. We show that the desired spin-motion entangled state can be created by engineering time-dependent detuning and spin-motion coupling which allows one to adiabatically transfer phonon excitations into collective spin excitations. Such a time-dependent control of the spin-phonon interaction is widely used for the creation of entangled spin and motion states \cite{Linington2008,Linington2008_1,Hume2009,Toyoda2011,Lechner2016,Kirkova2021}. We show that the super-Heisenberg scaling of the parameter estimation precision can be observed even for a small number of ions relevant to near-term experiments. Our adiabatic state preparation technique is insensitive to the motional mode coupled to the spins and can be applied outside the Lamb-Dicke regime. Furthermore, we show that the metrological gain obtained with the use of the proposed spin-motion state is preserved even in the presence of spin dephasing.

\emph{Model.-} 
We consider an ion chain with $N$ ions confined in a Paul trap. The coupling between the spin and motional degrees of freedom can be realized by driving a red sideband transition, described by the Tavis-Cummings Hamiltonian
\begin{equation}
\hat{H}_{\rm TC} (t) = \Delta (t) \hat{S}_z +\frac{\lambda (t)}{\sqrt{N}} (\hat{S}^{+}\hat{a} + \hat{S}^{-}\hat{a}^{\dagger}). \label{TC}
\end{equation}
Here $\Delta(t)$ is the time-dependent detuning to the center-of-mass red sideband resonance and $\lambda(t)$ is the homogeneous time-dependent spin-motion coupling.  Here $\hat{a}^{\dag}$ and $\hat{a}$ are the creation and annihilation operators of the collective center-of-mass phonons, $\hat{S}_{z}=\frac{1}{2}\sum_{l=1}^{N}\hat{\sigma}^{z}_{l}$ and $\hat{S}^{+}=\sum_{l=1}^{N}\sigma^{+}_{l}$ ($\hat{S}^{-}=(\hat{S}^{+})^{\dag}$) are the collective spin operators with $\hat{\sigma}^{z}_{l}$ being the Pauli operator for the $l$th spin and respectively $\hat{\sigma}^{+}_{l}$ is the spin raising operator.

The TC Hamiltonian commutes with the excitation number operator $\hat{N}=\hat{a}^{\dag}\hat{a}+\hat{S}_{z}+S$, and therefore the total number of collective spin and phonon excitations is conserved. Consequently, the full Hilbert space is decomposed into the subspaces with well-defined number of excitations $N=n_{s}+n$ where $n_{s}=0,1,\ldots,2S$ is the number of collective spin excitations with $S=N/2$ being the maximal spin length and respectively $n$ is the phonon number. 

\emph{Adiabatic Transition.-} 
The creation of the highly entangled spin-motion state begins by preparing the system initially in the product state $\left|\psi(t_{i})\right\rangle=\left|S,-S\right\rangle\left|\zeta\right\rangle$ where $|S,m\rangle$ is the eigenstate of $\hat{S}_{z}|S,m\rangle=m|S,m\rangle$ ($m=-S\ldots S$) and $\left|\zeta\right\rangle=\hat{S}(\zeta)\left|0\right\rangle$ is the single mode squeezed state, which is described by the unitary squeeze operator $\hat{S}(\zeta)=e^{\frac{1}{2}(\zeta^{*}\hat{a}^{2}-\zeta \hat{a}^{\dag 2})}$ where $\zeta=r e^{i\phi}$ is the squeezing parameter with amplitude $r$ and phase $\phi$ and $|n\rangle$ ($n=0,1,\ldots$) is the Fock state of the quantum harmonic oscillator. It is convenient to express the squeezed state in the Fock basis such that we have $|\zeta\rangle=\sum_{n=0}^{\infty}a_{2n}|2n\rangle$  where $a_{2n}=(\sqrt{(2n)!}/2^{n}n!)(\tanh^{n}(r)/\sqrt{\cosh(r)})e^{-i n\phi}$. Therefore, the initial state can be rewritten as $|\psi(t_{i})\rangle=\sum_{n=0}a_{2n}|\psi_{2n}(t_{i})\rangle$ where $|\psi_{2n}(t_{i})\rangle=\left|S,-S\right\rangle\left|2n\right\rangle$.
\begin{figure}
\includegraphics[width=0.57 \textwidth]{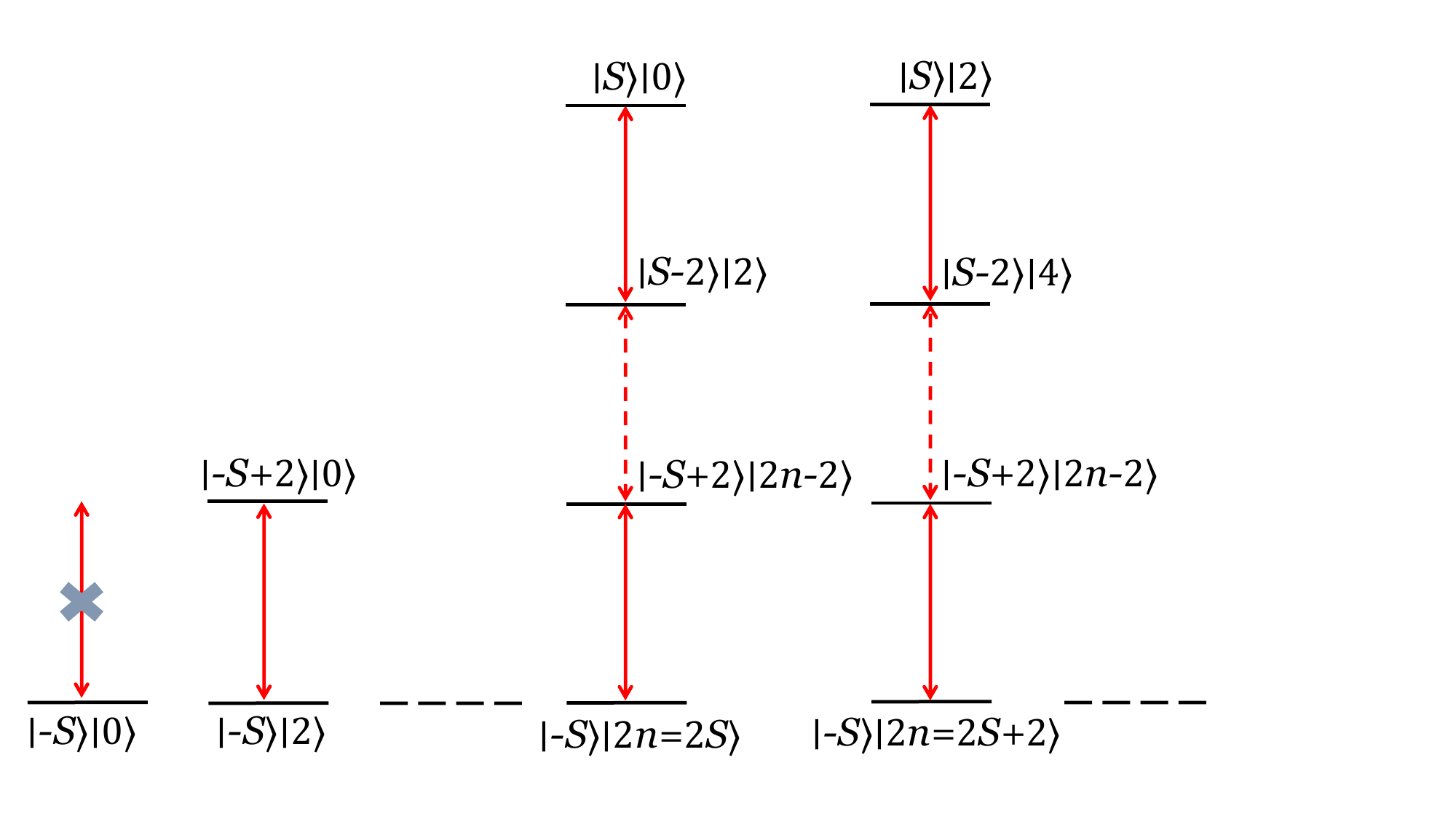}
\caption{Linkage pattern of the collective states of a string of ions driven by red-sideband laser. Spins are initially prepared in their electronic ground state $\left|S,-S\right\rangle=\left|\downarrow\downarrow\ldots\downarrow\right\rangle$ and the vibration center-of-mass mode is in squeezed state $|\zeta\rangle=\sum_{n=0}^{\infty}a_{2n}|2n\rangle$.}
\label{fig1}
\end{figure}

We assume that at the initial moment the detuning $\Delta(t_{i})$ is much larger than the spin-motion coupling $\lambda(t_{i})$, such that $|\Delta(t_{i})|\gg\lambda(t_{i})$ and $\Delta(t_{i})<0$. This implies that each of the states $|\psi_{2n}(t_{i})\rangle$ is an eigenstate of the TC Hamiltonian (\ref{TC}), namely $\hat{H}_{\rm TC}(t_{i})|\psi_{2n}(t_{i})\rangle=-S\Delta(t_{i})|\psi_{2n}(t_{i})\rangle$. Then we adiabatically vary the detuning $\Delta(t)$, so that we end up with $\Delta(t_{f})\gg\lambda(t_{f})$ and $\Delta(t_{f})>0$. As long as the adiabaticity is maintained, the system remains in the same eigenstates of the TC Hamiltonian (\ref{TC}) at all times. Since the number of excitations is preserved, each of the eigenstates $|\psi_{2n}(t_{i})\rangle$ is adiabatically transformed into the final eigenstate $|\psi_{2n}(t_{f})\rangle=|S,-S+2n\rangle|0\rangle$ for $0\leq 2n\leq 2S$ (we assume even number of ions) indicating that the phonon excitations are transferred into collective spin excitations, as is shown in Fig. \ref{fig1}. Because the maximum number of spin excitation is $n_{s}=2S$ where all spins are in the excited states the initial state $|\psi_{2n}(t_{i})\rangle$ with phonon number $2n>2S$ adiabatically evolves into $|\psi_{2n}(t_{f})\rangle=|S,S\rangle|2n-2S\rangle$. Therefore, the initial state with all spins in the ground state and phonon squeezed state is adiabatically transformed into the state
\begin{equation}
|\psi\rangle = \sum_{n=0}^{S} c_{2n}|S,-S+2n\rangle |0\rangle + \sum_{n=1}^{\infty} c_{2n+2S} |S,S\rangle |2n \rangle,\label{pi}
\end{equation}
where the probability amplitudes are $c_{2n}=a_{2n}e^{-i\epsilon_{n_{s},n}}$ and $\epsilon_{n_{s},n}=\int_{t_{i}}^{t_{f}}E_{n_{s},n}(t)dt$ are the adiabatic phases which appear due to the adiabatic evolution.

\emph{Quantum Metrology.-} 
Let us now consider the capability for high precision quantum metrology based on the quantum state (\ref{pi}). We assume that an unknown phase $\varphi$, the parameter we wish to estimate, is imprinted by a collective rotation about the $\hat{z}$ axis, described by $\hat{R}_{z}=e^{-i\varphi \hat{S}_{z}}$. Hence, the state vector becomes $|\psi^{\prime}\rangle=\hat{R}_{z}|\psi\rangle$. We can evaluate the QFI, which is a quantity of a central importance in quantum metrology as it is related to the fundamental precision of measurement uncertainty, by means of the quantum Cram\'er-Rao bound $\delta\varphi\geq 1/\sqrt{\mathcal{F}_{\varphi}}$, where $\mathcal{F}_{\varphi}$ is the QFI associated with parameter $\varphi$ \cite{Paris2009,Pezze2014,Liu2020}. For a pure state $|\psi^{\prime}\rangle$ we have $\mathcal{F}_{\varphi}=4\Delta^{2}\hat{S}_{z}$, where $\Delta^{2}\hat{S}_{z}=\langle \hat{S}^{2}_{z}\rangle-\langle \hat{S}_{z}\rangle^{2}$ is the variance of $\hat{S}_{z}$ with respect to the state ($\ref{pi}$). Then, it is straightforward to show that the variance of $\hat{S}_{z}$ is given by
\begin{eqnarray}
\Delta^{2}\hat{S}_{z}&=&\sum_{n=0}^{S} |c_{2n}|^{2} (-S+2n)^2 +S^2\sum_{n=1}^{\infty} |c_{2n+2S}|^{2}\notag\\
&&-\left( \sum_{n=0}^{S}|c_{2n}|^{2} (-S+2n) +S\sum_{n=1}^{\infty} |c_{2n+2S}|^{2}\right)^2.\label{sum}
\end{eqnarray}
The sums in Eq. (\ref{sum}) can be exactly evaluated such that one can derive an explicit expression for the QFI. We obtain
\begin{eqnarray}
\mathcal{F}_{\varphi}&=&8 \bigg[\frac{z}{(1-z)^{2}}-\frac{2 z^{2b-1}(1-z)}{\pi}\Gamma^{2}(b){}^{}_{2}\tilde{F}^{2}_{1}(a,b,c;z)\notag\\
&&-\frac{z^{b-\frac{1}{2}} \; \Gamma(b)}{\sqrt{\pi(1-z)}}\big(2S(1-z)^{-\frac{1}{2}}{}^{}_{2}\tilde{F}^{}_{1}\left(\frac{a}{4},b-\frac{1}{2},c;z\right)\notag\\
&&+(2+2S(1-z)-z){}^{}_{2}\tilde{F}_{1}^{}(a,b,c;z)\big) \bigg]. \label{QFI_Final}
\end{eqnarray} 
\begin{figure}
\includegraphics[width=0.51 \textwidth, center]{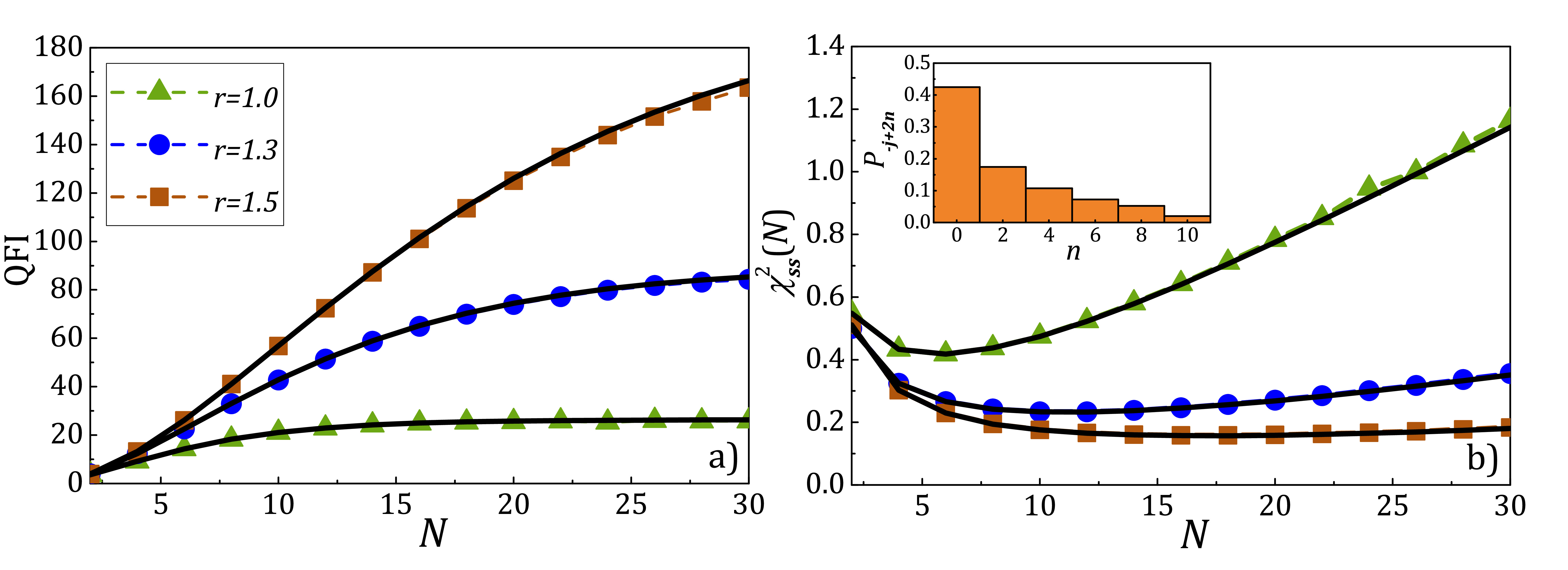}
\caption{(a) Quantum Fisher information as a function of the number of spins $N$. We assume that the system is prepared initially in the state $|\psi(t_{i})\rangle = \left|S,-S\right\rangle|\zeta\rangle$. We numerically evaluate $\mathcal{F}_{\varphi}=4\Delta^{2}\hat{S}_{z}$ and compare it with the analytical expression (\ref{QFI_Final}) (solid lines) for various squeezing amplitudes $r$. (b) $\chi^2_{\rm ss}$ as a function of $N$. (inset) A probability histogram for $P_{-S+2n}=|\langle\psi_{2n}(t_{f})|\psi\rangle |^2 \approx |c_{2n}|^2$ for $r=1.5$ and $N=10$. The detuning and the spin-phonon coupling are $\Delta(t)=\Delta_{0}\sin(\gamma t/2)$, $\lambda(t)=\lambda_{0}\cos^{2}(\gamma t/2)$ and the interaction time varies $t\in [t_{i},t_{f}]$ with $t_{i}=-t_{f}$, $t_{f}=\pi/\gamma$. The parameters are set to $\Delta_0/2 \pi  = 45$ kHz,\;$\lambda_0/2 \pi  = 20$ kHz,\;$ \gamma/ 2 \pi = 1$ kHz. }
\label{fig2}
\end{figure}Here $z=\tanh^{2}(r)$, $\Gamma(b)$ is the Euler's gamma function, and ${}^{}_{2}\tilde{F}_{1}^{}(a,b,c;z)$ is the regularized hypergeometric function with $a=2$, $b=\frac{3}{2}+S$, and $c=2+S$. 

We now examine two regimes depending on the initial phonon squeezing. In Fig. \ref{fig2}(a) we show the QFI in the weak squeezing regime which is characterized with $z<1$. The QFI increases with the number of spins and reaches a level of saturation for $N\gg1$ where it becomes independent of $N$ and is given by $\mathcal{F}^{(N\gg 1)}_{\varphi}=2\sinh^{2}(2r)$. Note that by increasing the squeezing amplitude $r$ the saturation level is reached for a larger number of spins, as is shown in Fig. \ref{fig2}(a). In order to examine the spin-squeezing and respectively the enhancement of the parameter estimation over the SNL we show in Fig. \ref{fig2}(b) the ratio $\chi^{2}_{\rm ss}(N)=N/\mathcal{F}_{\varphi}$ \cite{Pezze2009}. The parameter $\chi^{2}_{\rm ss}$ quantifies the multiparticle entanglement in the state (\ref{pi}) and provides a necessary and sufficient condition to achieve a sub shot-noise sensitivity of the parameter estimation as long as $\chi^{2}_{\rm ss}<1$. We see that in the regime of weak squeezing, only a small number of probability amplitudes $c_{2n}$ around $n=0$ contribute to the expression (\ref{pi}), see Fig. \ref{fig2}(b) (inset). Therefore, the second term in the quantum state (\ref{pi}) makes no contribution and the adiabatic process completely transfers the phonon excitations into collective spin excitations. The resulting state is a coherent superposition between the collective spin states $|S,m\rangle$ which is disentangled with the quantum harmonic oscillator. The parameter is $\chi^{2}_{\rm ss}<1$ and decreases with the number of spins $N$ up to some optimal value as is shown in Fig. \ref{fig2}(b). For example the numerical fit of the parameter $\chi^{2}_{\rm ss}$ for $r=1$ and $r=1.5$ show that it decreases with $N$ as $\chi^{2}_{\rm ss}\propto N^{-0.2}$ and $\chi^{2}_{\rm ss}\propto N^{-0.3}$ leading to sub shot-noise sensitivity. For a large $N$ and fixed $r$ the QFI tends to $\mathcal{F}_{\varphi}\rightarrow \mathcal{F}^{(N\gg 1)}_{\varphi}$ and respectively the parameter $\chi^{2}_{\rm ss}$ begins to increase linearly with $N$.
\begin{figure}
\includegraphics[width=0.55 \textwidth, center]{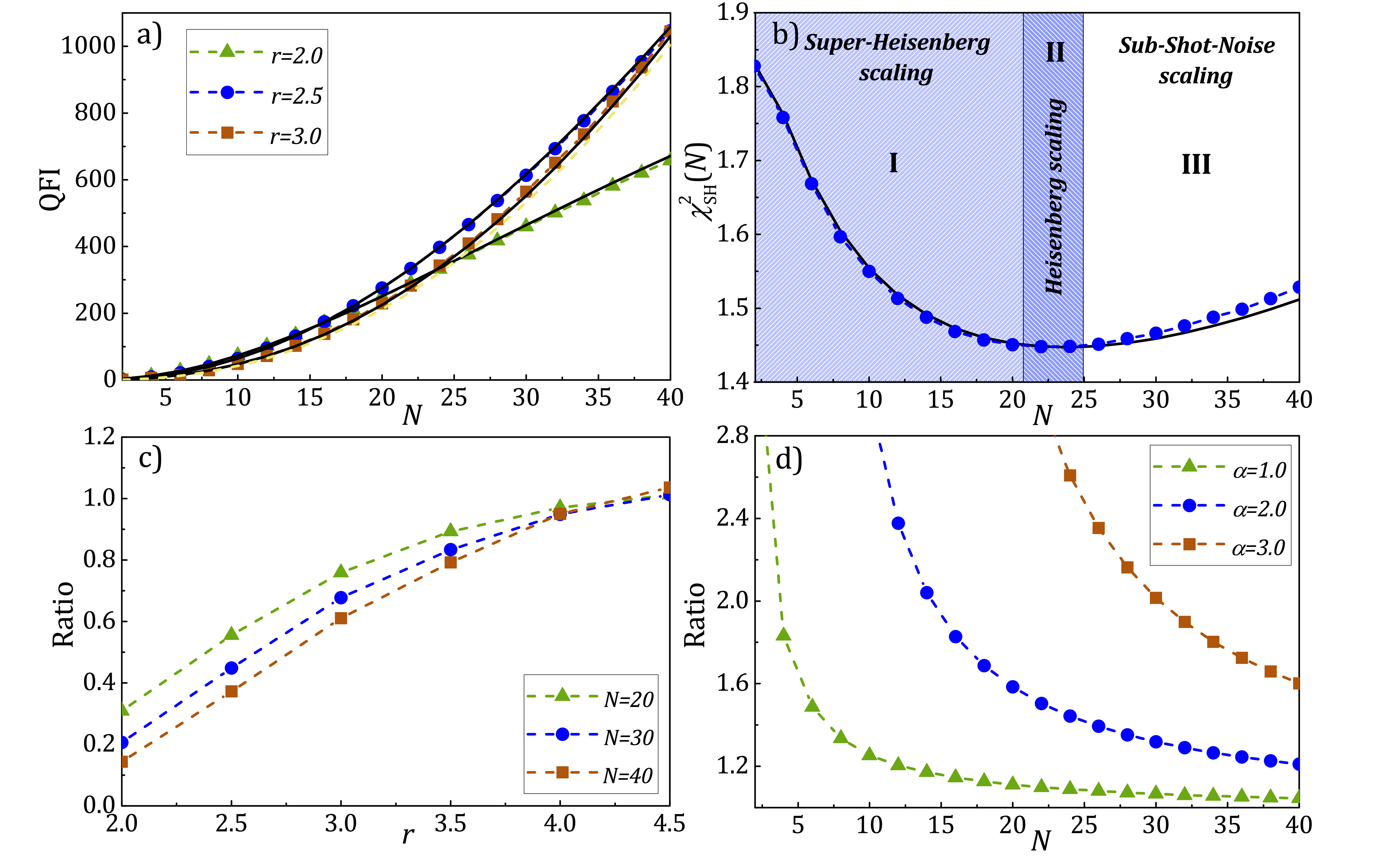}
\caption{(a) Quantum Fisher information as a function of the number of spins $N$. We numerically evaluate $\mathcal{F}_{\varphi}=4\Delta^{2}\hat{S}_{z}$ and compare it with the analytical expression (\ref{QFI_Final}) (solid lines) for various squeezing amplitudes $r$ and with the asymptotic formula (\ref{ass}) for $r=3.0$ (yellow dashed line). (b) $\chi^{2}_{\rm SH}$ as a function of $N$ for $r=2.5$. The three regions with different scalings are separated by vertical solid lines. (c) Ratio of the exact result for the QFI to the asymptotic expression $\mathcal{F}^{\rm max}_{\varphi}$ for different number of spins $N$.  (d) Ratio of the QFI to the QFI for a displaced squeezed initial state $|\psi_{(D)}(t_{i})\rangle =\hat{D}(\alpha)|\psi(t_{i})\rangle$, for various displacement amplitudes and $r=2.5$.}
\label{fig3}
\end{figure}

Remarkably, by increasing the phonon squeezing amplitudes one can reach the Heisenberg scaling of the QFI and even overcome it. We define a strong phonon squeezing regime according to $z\rightarrow 1$. Then the asymptotic behaviour of the QFI (\ref{QFI_Final}) becomes
\begin{eqnarray}
\mathcal{F}_{\varphi}&\sim&\frac{16}{315} \bigg[\frac{336\sqrt{1-z}}{\sqrt{\pi}}\left(\frac{N}{2}\right)^{\frac{5}{2}}-\frac{560(1-z)}{\pi}\left(\frac{N}{2}\right)^{3}\notag\\
&&-\frac{48(1-z)^{\frac{3}{2}}}{\sqrt{\pi}}\left(\frac{N}{2}\right)^{\frac{7}{2}}+\frac{224(1-z)^{2}}{\pi}\left(\frac{N}{2}\right)^{4} \bigg].\label{ass}
\end{eqnarray}
Therefore, up to the leading order we find a super-Heisenberg scaling law of the QFI as $\mathcal{F}_{\varphi}\propto N^{5/2}$. The underlying mechanism that leads to the super-Heisenberg scaling is the creation of a strongly entangled spin-motion state. As the initial phonon squeezing increases, the second term in expression (\ref{pi}) becomes more significant, because for a given number of spins only a finite number of phonons can be adiabatically transferred into collective spin excitations. Hence in the limit of strong phonon squeezing the maximally excited spin collective state $|S,S\rangle$ becomes coupled to a large number of Fock states, which leads to an entangled spin-motion state. As a figure of merit for super-Heisenberg scaling of the parameter estimation precision we introduce the measure $\chi^{2}_{\rm SH}=N^{2}/\mathcal{F}_{\varphi}$. As long as $\chi^{2}_{\rm SH}$ decreases with $N$ the scaling law of the QFI is $\mathcal{F}_{\varphi}\propto N^{l}$ with $l>2$. In order to overcome the Heisenberg limit of precision, we also require $\chi^{2}_{\rm SH}<1$. 
In Fig. \ref{fig3}(a) we show the QFI in the strong phonon squeezing regime. We find very good agreement between the exact result with Hamiltonian (\ref{TC}) and the analytical expression (\ref{QFI_Final}). We also compare the exact result of the QFI with the asymptotic formula (\ref{ass}) which is very accurate in the limit of $z\rightarrow1$. In Fig. \ref{fig3}(b) we show the parameter $\chi^{2}_{\rm SH}$ versus the number of spins $N$. We distinguish three regions with different scaling laws for the QFI. In the first region (I) the QFI exhibits a super-Heisenberg scaling law. By increasing $N$ the probability amplitudes $c_{2n+2S}$ in (\ref{pi}) become smaller which weakens the spin-motional entanglement. In this intermediate region (II) the parameter $\chi^{2}_{\rm SH}$ is nearly constant which indicates that the QFI changes the scaling law to $\mathcal{F}_{\varphi}\propto N^{2}$. For a even larger number of spins the second term in (\ref{pi}) becomes negligibly small and the parameter estimation obeys a sub-shot noise limit of precision in region (III). We also emphasize that by increasing the phonon squeezing amplitude $r$, the region of $N$ in which the parameter $\chi^{2}_{\rm SH}$ decreases becomes larger. As can be seen from Eq. (\ref{ass}) the maximal super-Heisenberg scaling of the QFI is achieved in the limit $z\rightarrow 1$ where the leading term is given by $\mathcal{F}^{\rm max}_{\varphi}=\frac{256\sqrt{1-z}}{15\sqrt{\pi}}\left(\frac{N}{2}\right)^{\frac{5}{2}}$. In Fig. \ref{fig3}(c) we show the ratio $\mathcal{F}_{\varphi}/\mathcal{F}^{\rm max}_{\varphi}$, where $\mathcal{F}_{\varphi}$ is the exact result for QFI with Hamiltonian (\ref{TC}). For a squeezing amplitude approximately equal to $r\geq 4$ the asymptotic expression $\mathcal{F}^{\rm max}_{\varphi}$ very closely describes the exact behaviour of the QFI, and in this limit the phase sensitivity becomes $\delta\varphi=\gamma N^{-5/4}$ where $\gamma\approx 0.77(1-z)^{-1/2}$. Since the factor $\gamma$ increases with $r$ the phase sensitivity is always bounded by the Heisenberg limit in a sense that $\chi^{2}_{\rm SH}>1$. Indeed, increasing $N$ the higher order terms in (\ref{ass}) become significant which spoil the phase precision. Hence, in order to reach the asymptotic limit $\mathcal{F}^{\rm max}_{\varphi}$ we need to increase further $r$ which limits the phase sensitivity.

Our adiabatic approach to create a highly entangled spin-motion state can also be applied for different initial motional states. Consider an initial state $\left|\psi(t_{i})\right\rangle=\left|S,-S\right\rangle\left|\alpha,\zeta\right\rangle$, where $\left|\alpha,\zeta\right\rangle=\hat{D}(\alpha)\hat{S}(\zeta)\left|0\right\rangle$ is a displaced squeezed state with displacement operator $\hat{D}(\alpha)=e^{\alpha \hat{a}^{\dag}-\alpha^{*}\hat{a}}$ and amplitude $\alpha$. The adiabatic transition creates an entangled spin-motion state similar to (\ref{pi}) \cite{sup}. In Fig. \ref{fig3}(d) we show the ratio $\mathcal{F}^{\alpha=0}_{\varphi}/\mathcal{F}^{\alpha\neq 0}_{\varphi}$ for various $\alpha$. Remarkably, the ratio decreases with $N$, which indicates that the corresponding QFI scales even faster with $N$ compared to the case with $\alpha=0$. 

Finally, we discuss the spin squeezing parameter which quantifies the sensitivity enhancement for Ramsey spectroscopic measurements. The metrological gain can be evaluated by the the spin squeezing parameter \cite{Wineland1992}
\begin{equation}
\xi^2 = \frac {N\Delta^{2} \hat{S}_{\Vec{n}_{\perp}}}{|\langle \hat{\Vec{S}} \rangle|^2},\label{sp}
\end{equation}
where $\Vec{n}_{\perp}$ is a unit vector, perpendicular to the mean spin direction (MSD).
\begin{figure}
\includegraphics[width=0.50 \textwidth, center]{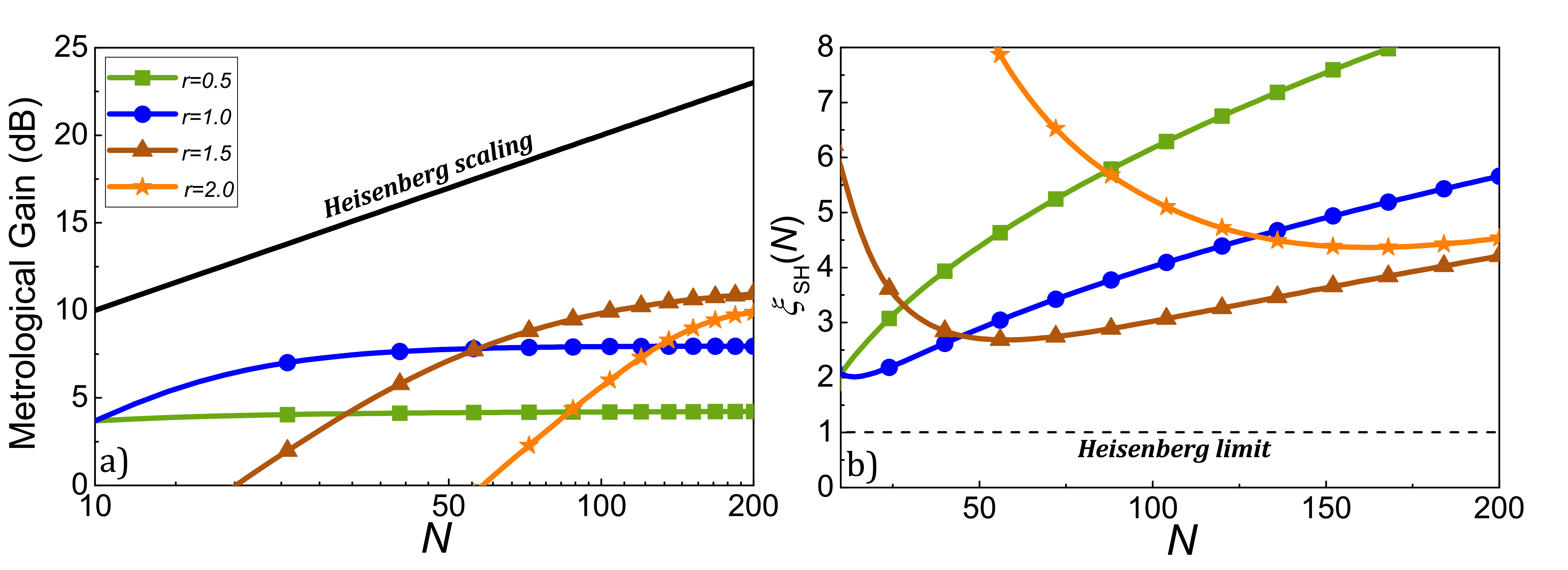}
\caption{(a) The metrological gain $[10\log_{10} (\delta\varphi_{\rm CSS}/\delta\varphi)^{2}]$ as a function of the number of spins $N$ for different squeezing amplitudes $r$. (b) The parameter $\xi_{\rm SH}$ as a function of $N$ for different $r$. The parameters used are set to $\Delta_0/2 \pi  = 2.5$ kHz,\;$\lambda_0/2 \pi  = 1.5$ kHz,\;$ \gamma/ 2 \pi = 5.5$ kHz. }
\label{fig4}
\end{figure}
The squeezing parameter (\ref{sp}) is the ratio $\xi^{2}=\delta\varphi^{2}/\delta\varphi^{2}_{\rm CSS}$ of the phase sensitivity of a general state versus the phase sensitivity using uncorrelated coherent spin state (CSS). As long as $\xi^{2}<1$ the state is a spin squeezed state and the phase sensitivity overcomes the SNL. In Fig. \ref{fig4}(a) we show the exact 
result for the metrological gain, where the MSD is along the $z$-direction and $\hat{S}_{\Vec{n}_{\perp}}=\cos(\alpha)\hat{S}_{x}+\sin(\alpha)\hat{S}_{y}$, where $\alpha$ is determined by the condition that $\Delta^{2} \hat{S}_{\Vec{n}_{\perp}}$ is minimal. We observe that for large amplitude $r$ the metrological gain becomes positive for large number of ions and its slope is larger than the slope of the Heisenberg scaling $\delta\varphi^{2}_{\rm H}=N^{-2}$ indicating that the phase sensitivity exhibits super-Heisenberg scaling up to a certain factor $\delta\varphi^{2} = c N^{-l}$, where $l>2$. In order to show this, in Fig. \ref{fig4}(b) we sketch the quantity $\xi_{\rm SH}^2=N\xi^{2}$ which is essentially the ratio $\delta\varphi^{2}/\delta\varphi^{2}_{\rm H}$. On one hand we see that $\xi_{\rm SH}>1$ indicating that the phase sensitivity is bounded by the Heisenberg limit. On the other hand $\xi_{\rm SH}$ decreases with $N$, pointing out that it decays faster than $N^{-2}$. For smaller $r$ we observe an almost complete transfer of the initial phonon squeezing into spin squeezing, namely that $\xi^{2}\approx e^{-2r}$ which leads to a sub shot-noise sensitivity of the phase estimation as is shown in  Fig. \ref{fig4}(a). Consider for example $N=52$ ions and $r=1.5$ we find 7.3 dB in the metrological gain which is approximately twice larger than the metrological gain experimentally observed in \cite{Franke2023}.

\emph{Summary and outlook.-} 
To conclude, we have proposed an entangled spin-motion state between a collective spin system and a vibrational mode for high precision quantum metrology which leads to a faster scaling of the parameter estimation uncertainty with the number of spins compared with the Heisenberg limit. The essence of our preparation technique is the adiabatic evolution which transfers the initial motional squeezing into a collective spin squeezing. For a strong motional squeezing a finite set of collective spin states becomes coupled to a large set of harmonic oscillator Fock states. We show that the spin-motion entanglement improves the phase sensitivity which can exhibit a super-Heisenberg scaling. Furthermore, our adiabatic technique can be applied even in the case of spatially inhomogeneous spin-motion coupling \cite{sup}. This is an important advantage over the GHZ states which are extremely susceptible to inhomogeneities induced by coherent non-collective interactions. The observation of the spin squeezing and the super-Heisenberg scaling can be achieved even in the presence of collective spin dephasing caused by magnetic field fluctuations and laser noise \cite{sup}. Our preparation technique can be further extent to arrays in 2D and 3D which provide diversity of normal modes for preparing complex spin-motion states.

\emph{Acknowledgments.-} We acknowledge the Bulgarian national plan for recovery and resilience, contract BG-RRP-2.004-0008-C01 (SUMMIT: Sofia University Marking Momentum for Innovation and Technological Transfer), project number 3.1.4. We thank Kilian Singer for useful discussions.

\clearpage
\pagebreak
\widetext
\begin{center}
\textbf{\large Super-Heisenberg scaling of the quantum Fisher information using spin-motion states 
}
\end{center}
\begin{center}
Venelin P. Pavlov and Peter A. Ivanov
\end{center}
\begin{center}
\emph{Center for Quantum Technologies, Department of Physics, St. Kliment Ohridski University of Sofia, James Bourchier 5 blvd, 1164 Sofia, Bulgaria}
\end{center}

\setcounter{equation}{0}
\setcounter{figure}{0}
\setcounter{table}{0}
\setcounter{page}{1}
\makeatletter
\renewcommand{\theequation}{S\arabic{equation}}
\renewcommand{\thefigure}{S\arabic{figure}}
\renewcommand{\bibnumfmt}[1]{[S#1]}
\renewcommand{\citenumfont}[1]{S#1}


\begin{section}{Spatially inhomogeneous spin-motion couplings}\label{QFI}
Here we extend the proposed adiabatic technique to the case of spatially inhomogeneous spin-motion coupling. Without loss of generality, we assume that the ion chain is addressed by laser field with frequency close to the breathing mode. The Hamiltonian is given by
\begin{equation}
\hat{H}_{\rm TC,B}=\frac{\Delta(t)}{2}\sum_{k=1}^{N}\sigma^{z}_{k}+\sum_{k=1}^{N}\lambda_{k}(t)(\sigma^{+}_{k}\hat{b}+\sigma^{-}_{k}\hat{b}^{\dag}),\label{TCB}
\end{equation}
$\hat{b}^{\dag}$ and $\hat{b}$ are the creation and the annihilation operators of phonon in the breathing mode. The spatially inhomogeneous spin-motion coupling is $\lambda_{k}(t)=\lambda(t)M_{k}$, where $M_{k}$ is the amplitude of the breathing mode on ion $k$. In that case we are not able to introduce collective spin operators and thus the numerical diagonalization requires to explore the full spin Hilbert space with dimension $2^{N}$. Although the coupling is spatially inhomogeneous, the Hamiltonian (\ref{TCB}) commutes with the total number excitation operator, $\hat{N}_{\rm B}=\hat{b}^{\dag}\hat{b}+\frac{1}{2}\sum_{k=1}^{N}\sigma^{z}_{k}$, such that $[\hat{N}_{\rm B},\hat{H}_{\rm TC,B}]=0$.

An interesting question arises whether this change in the interaction between the internal and the motional degrees of freedom alters the properties of the system in the strong bosonic squeezing regime compared with center-of-mass mode interaction. Again we assume that the system is prepared initially in the state with all spins in the electronic ground state and squeezed motion state of the breathing mode, namely $|\psi(t_{i})\rangle=\left|\downarrow\downarrow\ldots\downarrow\right\rangle|\zeta\rangle$.
Since the total number of excitations is preserved, the initial phonon excitation is adiabatically transformed into a spin excitations. The final entangled spin-motion state is 
\begin{equation}
|\psi\rangle=\sum_{n=0}^{S}c_{2n}|\psi_{-S+2n}\rangle|0\rangle+\sum_{n=1}^{\infty}c_{2n+2S}|\uparrow\uparrow\ldots\uparrow\rangle|2n\rangle.
\end{equation}
Here $|\psi_{-S+2n}\rangle$ is a spin state of $N$ ions sharing $-S+2n$ excitations. In Figs. \ref{figs1}(a) and (b) we investigate the parameters $\chi^{2}_{\rm ss}$ and $\chi^{2}_{\rm SH}$ respectively, where we can see that the behavior is practically the same as in the homogeneous coupling case, seeing that $\chi^{2}_{\rm ss}<1$ and $\chi^{2}_{\rm SH}$ decreases with the number of spins $N$, preserving the super-Heisenberg scaling of the QFI.

\begin{figure}[h]
\includegraphics[width=0.75 \textwidth, center]{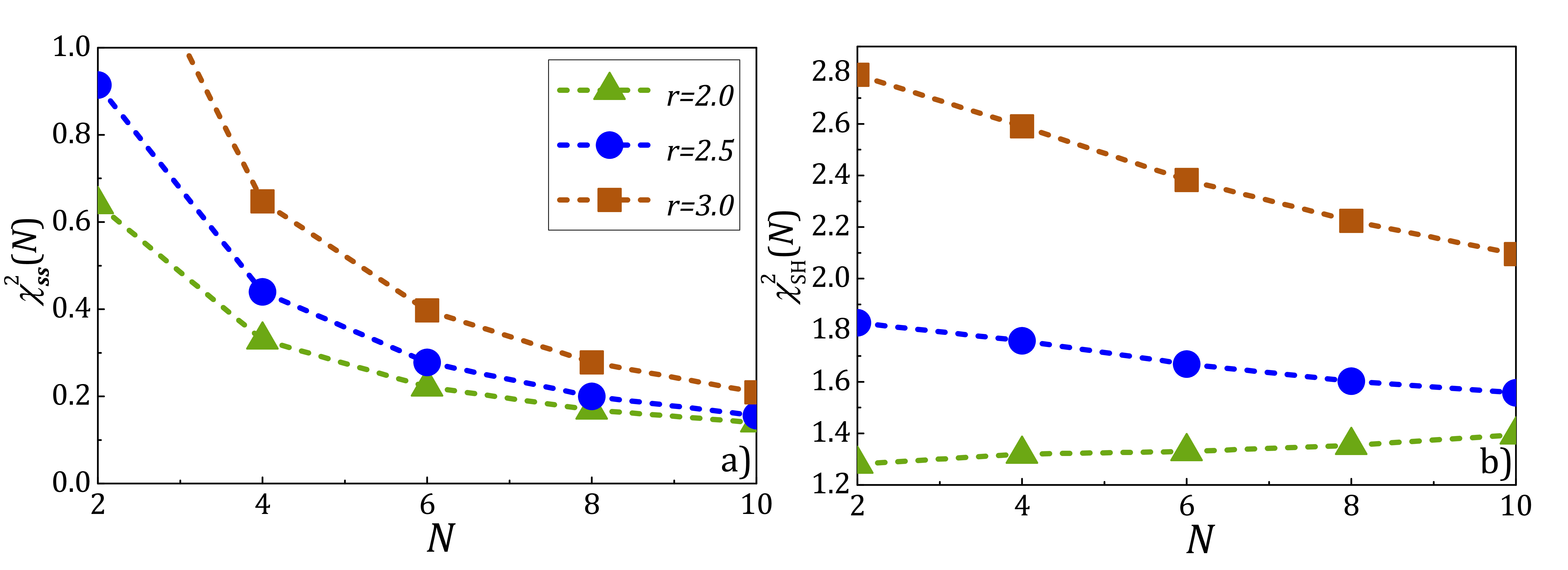}
\caption{(a) $\chi^{2}_{\rm ss}$ as a function of the number of spins $N$ for various squeezing amplitudes $r$. (b) $\chi^{2}_{\rm SH}$ as a function of the number of spins $N$. The parameters are set to $\Delta_0/2 \pi  = 45$ kHz,\;$\lambda_0/2 \pi  = 20$ kHz,\;$ \gamma/ 2 \pi = 1$ kHz.}
\label{figs1}
\end{figure}

\end{section}
\begin{section}{Collective Spin Dephasing}
We further expand our method and discuss the effect of a collective spin dephasing to the system. Such a collective dephasing noise appears due to magnetic field fluctuations and laser noise. For this purpose we use an open system formalism in terms of the master Lindblad equation $\partial_{t} \hat{\rho} = \mathcal{L} \hat{\rho}$, where $\mathcal{L}$ is the Liouvillian superoperator which describes the dynamics of the system. For the Tavis-Cummings Hamiltonian and collective spin dephasing the Lindblad equation reads
\begin{equation}
\partial_{t} \hat{\rho} = -\textit{i}[\hat{H}_{\rm TC}, \hat{\rho}] + \Gamma \mathcal{D}_{\hat{S}_{z}}[\hat{\rho}],
\end{equation}
where $\Gamma$ is the collective dephasing rate and $\mathcal{D}_{\hat{S}_{z}}[\hat{\rho}] = 2\hat{S}_{z} \hat{\rho} \hat{S}_{z} - \{\hat{S}^{2}_{z},\hat{\rho}\}$ is the Lindblad superoperator with quantum jump operator $\hat{S}_{z}$. To investigate how the metrological properties of the system change in the strong bosonic squeezing regime we again consider the parameters $\chi^{2}_{\rm ss}$ and $\chi^{2}_{\rm SH}$. In Fig. {\ref{figs2}}(a) we see that the increase of the dephasing rate for a fixed number of the spins $N$ saturates $\chi^{2}_{\rm ss}$ to a value that is still smaller than unity, indicating that the presence of a sub-shot sensitivity in the system is observed even in the presence of dephasing. Interestingly, in Fig. {\ref{figs2}}(b) we see that for a small dephasing rate the super-Heisenberg scaling is still present in the system, but vanishes when we increase the parameter, owing to the fact we no longer observe a decrease in the parameter $\chi^{2}_{\rm SH}$ with the number of spins $N$. 
\begin{figure}[h]
\includegraphics[width=0.75 \textwidth, center]{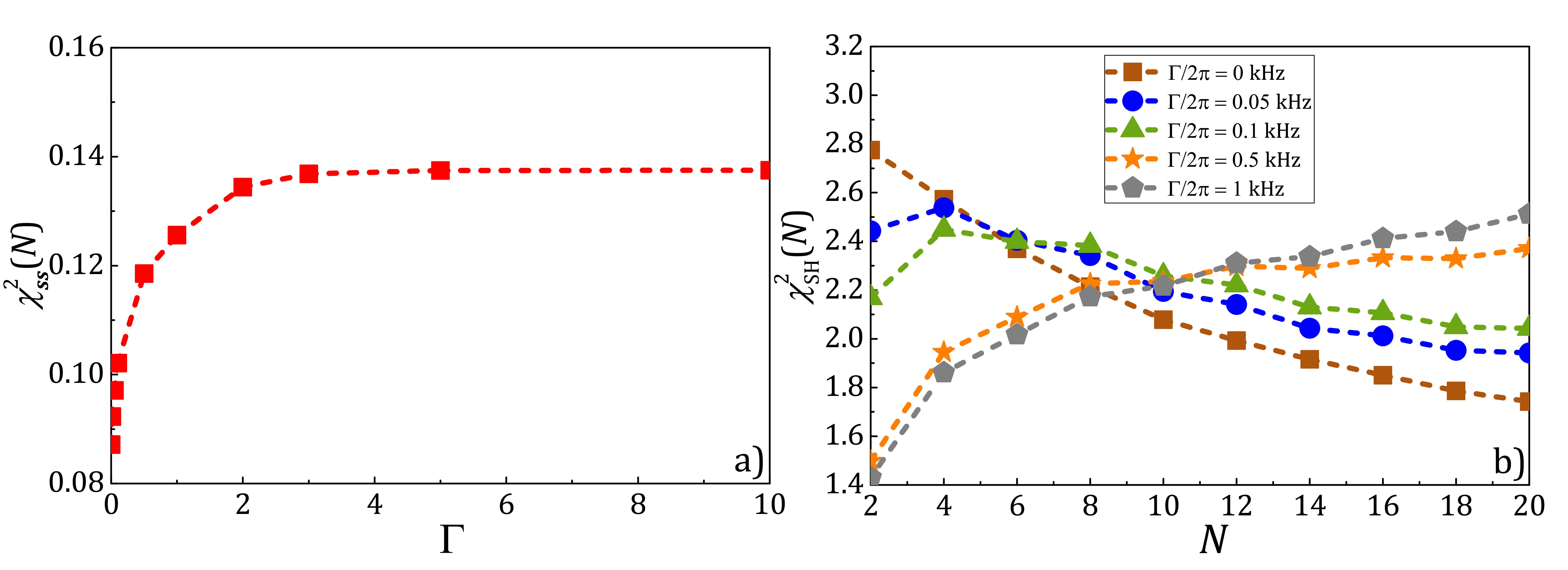}
\caption{(a) $\chi^{2}_{\rm ss}$ as a function of the collective dephasing rate $\Gamma$ for $r=3.0$ and $N=20$. (b) $\chi^{2}_{\rm SH}$ as a function of the number of spins $N$ for various dephasing rates $\Gamma$ and $r$=3.0 }
\label{figs2}
\end{figure}

\begin{section}{Displaced Squeezed Initial State}
Our adiabatic technique can be applied for a variety of initial motion states of the trapped ions. Here we consider the initial state $\left|\psi(t_{i})\right\rangle=\left|S,-S\right\rangle\left|\alpha,\zeta\right\rangle$, where $\left|\alpha,\zeta\right\rangle=\hat{D}(\alpha)\hat{S}(\zeta)\left|0\right\rangle$ is a displaced squeezed state. Here $\hat{D}(\alpha)=e^{\alpha \hat{a}^{\dag}-\alpha^{*}\hat{a}}$ is the displacement operator with amplitude $\alpha$. We express this state in the Fock basis $\left|\alpha,\zeta\right\rangle=\sum_{n=0}^{\infty}b_{n}|n\rangle$, where the probability amplitudes $b_n$ are given by
\begin{equation}
b_{n}=\frac{e^{-\frac{|\alpha|^2}{2}-\frac{\alpha^{*2}}{2}e^{i\phi} \tanh{r}}}{\sqrt{\cosh r}} 
\sum_{n=0}^{\infty} \frac{(e^{i\phi} \tanh{r})^{n/2}}{\sqrt{2^{n}n!}}H_n[\gamma(e^{i\phi} \sinh{2r})^{-1/2}],
\end{equation}
where $\gamma=\alpha \cosh r + \alpha^{*} e^{i \phi} \sinh r$, and $H_n[x]$ are the Hermite polynomials. In contrast to the case of $\alpha=0$ now the Fock state representation of $\left|\alpha,\zeta\right\rangle$ contains even and odd number of phonon excitations. Therefore, the adiabatic evolution transforms the initial state into the final state
\begin{equation}
\left|\psi(t_{f})\right\rangle = \sum_{n=0}^{2S} d_{n}|S,-S+n\rangle |0\rangle + \sum_{n=1}^{\infty} d_{n+2S} |S,S\rangle |n \rangle,\label{pi2}
\end{equation}
where the probability amplitudes are $d_{n}=b_{n}e^{-i\epsilon_{n_{s},n}}$ and $\epsilon_{n_{s},n}=\int_{t_{i}}^{t_{f}}E_{n_{s},n}(t)dt$ are the adiabatic phases which appear due to the adiabatic evolution. Using (\ref{pi2}), we find that the variance $\Delta^{2}\hat{S}_{z}$ is given by
\begin{equation}
\Delta^{2}\hat{S}_{z}=\sum_{n=0}^{2S} |d_{n}|^{2} (-S+n)^2 +S^2\sum_{n=1}^{\infty} |d_{n+2S}|^{2}
-\left( \sum_{n=0}^{2S}|d_{n}|^{2} (-S+n) +S\sum_{n=1}^{\infty} |d_{n+2S}|^{2}\right)^2.
\end{equation}
However, compared to Eq. (\ref{QFI_Final}), we are not able to derive an analytical expression for the QFI. Hence we numerically solve the time-dependent Scr\"odinger equation and plot the ratio $\mathcal{F}^{\alpha=0}_{\varphi}/\mathcal{F}^{\alpha\neq 0}_{\varphi}$ as is shown in Fig. \ref{fig3}(c) in the main text.

\end{section}
\end{section}
 

\begin{thebibliography}{99}

\bibitem{Degen2017} C. L. Degen, F. Reinhard, and P. Cappellaro, Rev. Mod. Phys. \textbf{89}, 035002 (2017).

\bibitem{Pezze2018} L. Pezz\'e, A. Smerzi, M. K. Oberthaler, R. Schmied, and P. Treutlein, Rev. Mod. Phys. \textbf{90}, 035005 (2018).

\bibitem{Giovannetti2011} V. Giovannetti, S. Lloyd, and L. Maccone, Nat. Photon. \textbf{5}, 222 (2011).

\bibitem{Giovannetti2006} V. Giovannetti, S. Lloyd, and L. Maccone, Phys. Rev. Lett. \textbf{96}, 010401 (2006).

\bibitem{Kitagawa1993} M. Kitagawa and M. Ueda, Phys. Rev. A \textbf{47}, 5138 (1993).

\bibitem{Ma2011} J. Ma, X. Wang, C. P. Sun, and F. Nori, Physics Reports, \textbf{509}, 89 (2011).

\bibitem{Comprain2022} T. Comprain, F. Mezzacapo, M. R.-de-Saint-Vincent, and T. Roscilde, Phys. Rev. Lett. \textbf{129}, 113201 (2022).

\bibitem{Comparin2022_1} T. Comparin, F. Mezzacapo, and T. Roscilde, Phys. Rev. Lett. \textbf{129}, 150503 (2022).

\bibitem{Swan2024} R. J. Lewis-Swan, J. C. Zuniga Castro, D. Barberena, and A. M. Rey, Phys. Rev. Lett. \textbf{132}, 163601 (2024).

\bibitem{Carrasco2024} S. C. Carrasco, M. H. Goerz, S. A. Malinovskaya, V. Vuletic, W. P. Schleich, and V. S. Malinovsky, Phys. Rev. Lett. \textbf{132}, 153603 (2024).

\bibitem{Feng2024} X. N. Feng, M. Zhang, and L. F. Wei, Phys. Rev. Lett. \textbf{132}, 220801 (2024).

\bibitem{Pavlov2023} V. P. Pavlov, D. Porras, and P. A. Ivanov, Phys. Scr. \textbf{98}, 095103 (2023).

\bibitem{Perlin2020} M. A. Perlin, C. Qu, and A. M. Rey, Phys. Rev. Lett. \textbf{125}, 223401 (2020).

\bibitem{Bonhet2016} J. G. Bonhet, B. C. Sawyer, J. W. Britton, M. L. Wall, A. M. Rey, M. Foss-Feig, and J. J. Bollinger, Science \textbf{352}, 1297 (2016).

\bibitem{Hosten2016} O. Hosten, N. J. Engelsen, R. Krishnakumar, and M. A. Kasevich, Nature \textbf{529}, 505 (2016).

\bibitem{Penafiel2020} E. Pedrozo-Panafiel, S. Colombo, C. Shu, A. F. Adiyatullin, Z. Li, E. Mendez, B. Braverman, A. Kawasaki, D. Akamatsu, Y. Xiao, and V. Vuletic, Nature \textbf{588}, 414 (2020).

\bibitem{Huang2023} M. Z. Huang, J. A. de la Paz, T. Mazzoni, K. Ott, P. Rosenbusch, A. Sinatra, C. L. G. Alzar, and J. Reicher, PRX Quantum \textbf{4}, 020322 (2023).

\bibitem{Bornet2023} G. Bornet, G. Emperauger, C. Chen, B. Ye, M. Block, M. Bintz, J. A. Boyd, D. Barredo, T. Comparin, F. Mezzacapo, T. Roscilde, T. Lahaye, N. Y. Yao, and A. Browaeys, Nature \textbf{621}, 728 (2023).

\bibitem{Franke2023} J. Franke, S. R. Muleady, R. Kaubruegger, F. Kranzl, R. Blatt, A. M. Rey, M. K. Joshi, and C. F. Roos, Nature \textbf{621}, 740 (2023).

\bibitem{Boixo2007} S. Boixo, S. T. Flammia, C. M. Caves, and JM Geremia, Phys. Rev. Lett. \textbf{98}, 090401 (2007).

\bibitem{Choi2008} S. Choi and B. Sundaram, Phys. Rev. A \textbf{77}, 053613 (2008).

\bibitem{Roy2008} S. M. Roy and S. L. Braunstein, Phys. Rev. Lett. \textbf{100}, 220501 (2008).

\bibitem{Napolitano2011} M. Napolitano, M. Koschorreck, B. Dubost, N. Behbood, R. J. Sewell, and M. W. Mitchell, Nature \textbf{471}, 486 (2011).

\bibitem{Zwierz2012} M. Zwierz, C. A. P\'erez-Delgado, and P. Kok, Phys. Rev. A \textbf{85}, 042112 (2012).

\bibitem{Rams2018} M. M. Rams, P. Sierant, O. Dutta, P. Horodecki, and J. Zakrzewski, Phys. Rev. X \textbf{8}, 021022 (2018).

\bibitem{Gietka2023} K. Gietka and H. Ritsch, Phys. Rev. Lett. \textbf{130}, 090802 (2023).

\bibitem{Lyu2020} C. Lyu, S. Choudhury, C. Lv, Y. Yan, and Q. Zhou, Phys. Rev. Research \textbf{2}, 033070 (2020).

\bibitem{Wineland1992} D. J. Wineland, J. J. Bollinger, W. M. Itano, F. L. Moore, and D. J. Heinzen, Phys. Rev. A \textbf{46}, R6797 (1992).

\bibitem{Wineland1998} D. J. Wineland, C. Monroe, W. M. Itano, D. Leibfried, B. E. King, and D. M. Meekhof, J. Res. Natl. Inst. Stand. Technol. \textbf{103}, 259 (1998).

\bibitem{Haffner2008} H. H\"affner, C. F. Roos, and R. Blatt, Phys. Rep. \textbf{469}, 155 (2008).

\bibitem{Schneider2012} C. Schneider, D. Porras, and T. Schaetz, Rep. Prog. Phys. \textbf{75}, 024401 (2012).

\bibitem{Monroe2021} C. Monroe, W. C. Campbell, L.-M. Duan, Z.-X. Gong, A. V. Gorshkov, P. W. Hess, R. Islam, K. Kim, N. M. Linke, G. Pagano, P. Richerme, C. Senko, and N. Y. Yao, Rev. Mod. Phys. \textbf{93}, 025001 (2021).

\bibitem{Linington2008} I. E. Linington and N. V. Vitanov, Phys. Rev. A \textbf{77}, 010302(R) (2008).

\bibitem{Linington2008_1} I. E. Linington, P. A. Ivanov, N. V. Vitanov, and M. B. Plenio, Phys. Rev. A \textbf{77}, 063837 (2008).

\bibitem{Hume2009} D. B. Hume, C. W. Chou, T. Rosenband, and D. J. Wineland, Phys. Rev. A \textbf{80}, 052302 (2009).

\bibitem{Toyoda2011} K. Toyoda, T. Watanabe, T. Kimura, S. Nomura, S. Haze, and S. Urabe, Phys. Rev. A \textbf{83}, 022315 (2011).

\bibitem{Lechner2016} R. Lechner, C. Maier, C. Hempel, P. Jurcevic, B. P. Lanyon, T. Monz, M. Brownnutt, R. Blatt, and C. F. Roos, Phys. Rev. A \textbf{93}, 053401 (2016).

\bibitem{Kirkova2021} A. V. Kirkova, W. Li, and P. A. Ivanov, Phys. Rev. Research \textbf{3}, 013244 (2021).

\bibitem{Paris2009} M. G. A. Paris, Int. J. Quantum Inf. \textbf{07}, 125 (2009).

\bibitem{Pezze2014} L. Pezz\'e and A. Smerzi, Atom Interferometry, in \emph{Proceedings of the International School of Physics "Enrico Fermi", Course 188, Varenna}, edited by G. M. Tino and M. A. Kasevich (IOS Press, Amsterdam, 2014), p. 691.

\bibitem{Liu2020} J. Liu, H. Yuan, X.-M. Lu, and X. Wang, J. Phys. A: Math. Theor. \textbf{53}, 023001 (2020).

\bibitem{Pezze2009} L. Pezz\'e and A. Smerzi, Phys. Rev. Lett. \textbf{102}, 100401 (2009).

\bibitem{sup} "See Supplemental Material at [URL will be inserted by publisher]".






\end{thebibliography}
\end{document}